# School Transport Electrification - Adoption, Strategies, Methods and Policy: A Comprehensive Review

**[1]Megh Bahadur KC, [2]Ziqi Song, Ph. D.***

*[1,2]Department of Civil, Structural and Environmental Engineering, University at Buffalo, 233 Ketter Hall, Buffalo, NY 14260*

Email: [1]meghbaha@buffalo.edu, [2]zqsong@buffalo.edu

Orcid ID: [1]0000-0001-5257-5779, [2]0000-0002-9693-3256

**Abstract:**

The move towards electric school buses (ESBs) marks a critical step in creating a healthier and more sustainable future for students. To meet the ambitious goal of zero-emission school buses by 2035, this review focuses on the need assessment, practices, gaps, challenges, and way forward. We conducted a comprehensive assessment of more than 100 relevant sources, resulting in a final investigation. In-depth, systematic, and qualitative content analysis with SWOT analysis produced critical insights into school transport electrification. The results showed that 1.8% of the total buses in the US have already been converted to electric, where California alone owning 29% of the buses. Subsidies from various agencies and programs have contributed to the rapid growth of electrification. However, challenges in cost, technology, and policies must be mitigated through innovation and stakeholder partnerships. Policy support is boosting subsidies, industry investment and market readiness. Equitable policy is important to support underserved and disadvantaged populations, which can be addressed through four key dimensions of equity: procedural, recognition, distributive, and reparative equity. Furthermore, the traditional bus deployment model is still the most common, whereas Transportation-as-a-Service (TaaS) is an innovative ESB deployment model with the potential to accelerate ESB adoption by integrating vehicle-to-grid. SWOT analysis indicated that the achievement of the zero-emission goal, autonomous driving, and repowered vehicle technology are the greatest opportunities. Dynamic electrification strategies, V2G technology and system resiliency are yet to be discovered, which could be crucial for mass electrification.

**Highlights:**

- Explored school bus electrification gaps and bridges the needs by identifying cost, technology and policy barriers
- Systematic in-depth content analysis and SWOT analysis have been performed
- Investments in infrastructure, battery and charger technological advancements play a vital role in accelerated school bus electrification
- Equitable electrification policy needs to be implemented, taking care of the underserved and disadvantaged population
- Dynamic strategies, vehicle-to-grid technology and system resiliency are yet to be discovered that have an immense impact on full-scale ESB electrification

| List of Abbreviations | | | |
|---|---|---|---|
| EBs | Electric Buses | | |
| ESBs | Electric School Buses | TCO | Total Cost of Ownership |
| V2G | Vehicle to Grid | CO2 | Carbon Dioxide |
| EVs | Electric Vehicles | NOx | Nitrogen Oxides |
| LCA | Lifecycle Cost Analysis | CO2e | Carbon Dioxide equivalent |
| EPA | Environmental Protection Agency | CNG | Compressed Natural Gas |
| BEBs | Battery Electric Buses | DGE | Diesel Gallon Equivalent |
| kWh | Kilowatt Hour | WRI | World Research Institute |
| HVIP | Hybrid and Zero-Emission Truck and Bus Voucher Incentive Program | NYSERDA | New York State Energy Research and Development Authority |
| USDOT | United States Department of Transportation | V2G | Vehicle-to-Grid |
| PM | Particulate Matter | GHG | Greenhouse Gas |
| NY | New York | SD | School District |

## 1. Introduction:

The US school transportation industry is the largest form of public transportation in terms of fleet size, as approximately 500,000 yellow buses transport 26 million pupils daily [1], [2]. In addition, New York has 45,000 school buses that transport 1.5 million students daily [3]. Among them, more than 95% of school buses run on high-polluting fossil fuels [4]. The transportation sector contributes nearly 30% of GHG emissions in the US and New York State [5]. These emissions compromise air quality and affect student health and academic performance. Furthermore, diesel exhaust causes respiratory diseases, such as asthma. Childhood asthma affects nearly 6.3 million children in the US [6]. Nationwide, diesel-powered school buses produce more than five million tons of emissions annually [7]. To combat the escalating climate crisis, the U.S. The EPA highlights the critical need to reduce almost all GHG emissions from the transportation sector by 2050 [8]. Therefore, it is essential to develop clean, safe, accessible, and equitable mobility systems to ensure a sustainable future. In addition, school transport decarbonization is becoming a national priority because of student health, environmental concerns, and global climate change.

In response to emission concerns, there is a movement towards transitioning conventional diesel fleets with alternative fuel bus technology such as BEBs, hydrogen fuel buses, Fuel cell vehicles and hybrid buses [9]. BEB are leading the transition towards zero-emission goals for several compelling reasons, such as lower operating costs with electricity as a fuel, easy installation of charging stations rather than establishing a hydrogen power plant, maturity in battery or charger technologies, well-to-wheel efficiency, and renewable energy synergy [10], [11], [12]. Furthermore, shifting the US yellow diesel buses to electric school buses led to a reduction of eight megatons in GHG emissions [13]. ESBs are expected to be 60% more fuel-efficient than their diesel counterparts [14]. Although the initial investment in electric buses may be higher, the long-term savings from reduced maintenance expenses can substantially offset these upfront costs, making electric school buses a financially and operationally efficient choice for school districts aiming to reduce their TCO [15]. The successful implementation of ESBs in Lake Shore Central School District, NY, demonstrated significant cost savings: a route that would cost $22 with a diesel bus costs only $4 when operated by an ESB, resulting in an estimated annual savings of approximately $15,000 [16].

However, several challenges exist in adopting ESBs, such as high upfront costs [17]. Additionally, the shift to zero-emission buses involves navigating a landscape filled with technological and policy uncertainties. Frequent changes in regulations, variability of incentives, and lack of established infrastructure can create a complex environment that may deter schools from making the transition [18], [19]. Moreover, the WRI's analysis report of 22 stakeholder organizations dedicated to equity and justice found that most stakeholder organizations lacked familiarity and involvement in electrifying school buses. They even found that they did not possess sufficient knowledge about recent ESB initiatives such as the Infrastructure Investment & Jobs Act or EPA's Clean School Bus Program [20]. These gaps highlight the necessity of an informed and equitable push towards school bus electrification. Furthermore, as school districts are new to electrification, they may have issues with the ESB deployment model. ESB deployment is crucial because it involves various components, including bus purchase, upkeep and maintenance, charger and charging station deployment, and demand energy and V2G management, which could be beyond the expertise of the school alone [21]. Therefore, we categorized and presented the ESB challenges into three main categories and their interrelationships, which are depicted in the Venn diagram below.

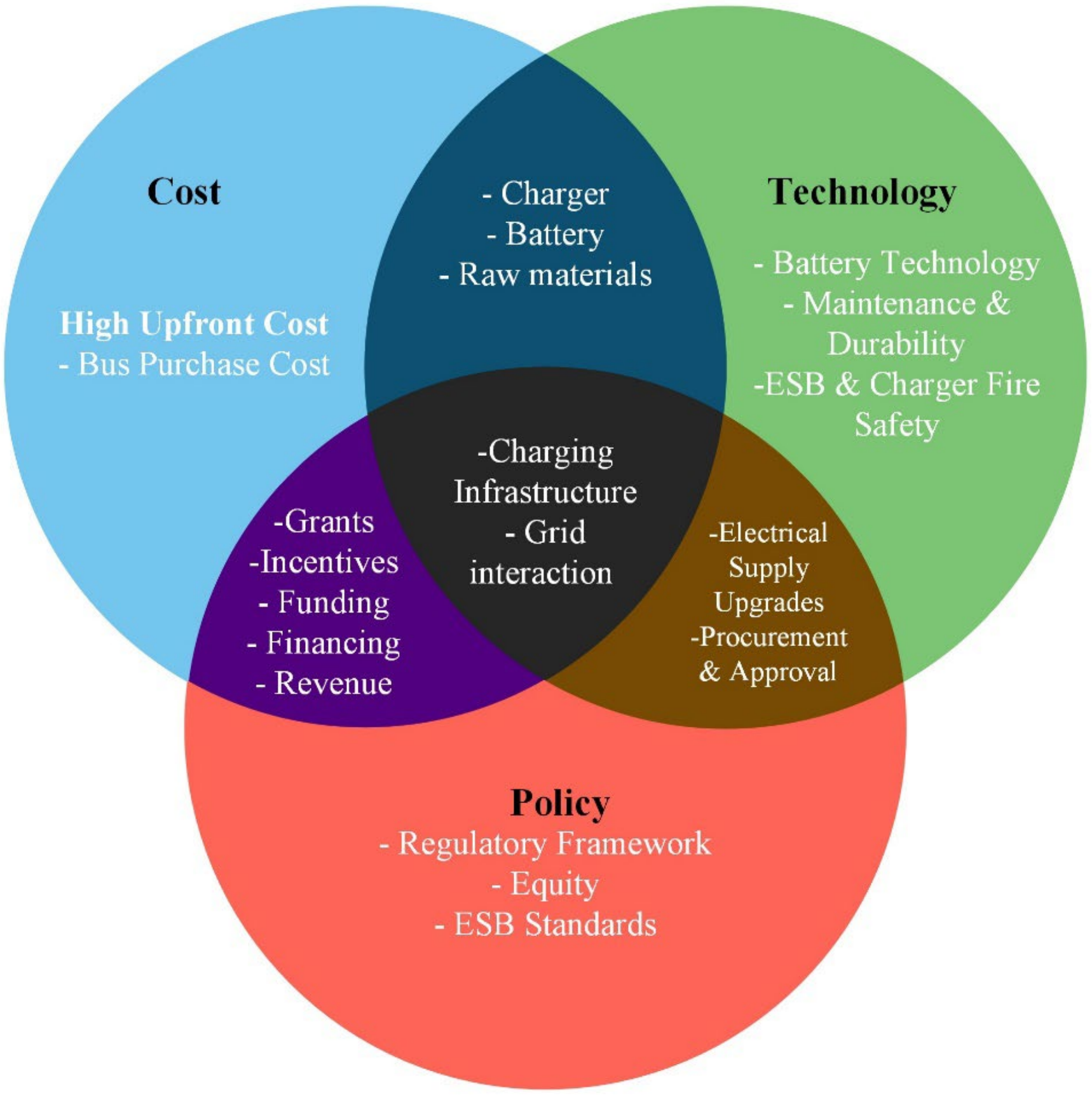

*Figure 1: ESB challenges and their interrelationship among the top categories*

Considering the interrelated aspects mentioned above, we need to integrate and synthesize the efforts made in ESB electrification and address the electrification challenges to mitigate the burning issue of pupils' health and transportation sustainability. Hence, this article provides a comprehensive literature review of ESB electrification and technology development to unveil the main research streams, shed light on research gaps, and provide information concerning methods for ESB deployment, tackle specific challenges, and inform policies, stakeholder participation, and research trends for the future. This article is developed around four questions to provide a thorough literature overview of the topic, which are presented below.

1. What are the primary drivers and barriers to the transition towards zero-emission school transportation, and how do these factors relate to different stakeholders, including electric bus manufacturers, bus operators, users, grid operators, and utility providers?
2. What are the current advancements and strategies supporting the widespread use of ESB in the United States? Additionally, which methodologies have proven effective in advancing ambitious sustainability and zero-emission objectives within the school transportation sector?
3. What policies and deployment models have been implemented to support school bus electrification, and how effective have these policies been in achieving progress?
4. What are the results of the SWOT analysis of school bus electrification, and what should be the prominent research directions in this field to ensure rapid electrification?

Our analytical methodology focused on integrating peer-reviewed, grey literature, white papers, and real-time data sources to map the most recent progress on ESB adoption and supporting policies. After finalizing the selected documents, we conducted an in-depth qualitative content analysis to systematically explore themes, patterns, and meanings within the topic of school transport electrification. Furthermore, we applied a SWOT analysis to deepen our understanding of the current state of ESB adoption, associated policies, and the direction of future research on large-scale school bus electrification. The results showed rapid progress in ESB adoption; for example, there were only 31 electric school buses on the road before 2012, but that number reached more than 3,750 in 2023 [22]. Most states that have developed statutory policies for zero-emission school buses aim to make school transportation zero-emission by 2035, such as California, New York, and Maine [18], [19]. Renewals, such as solar and wind energy integration policies, could further reduce the school's energy bills, along with a reduction in GHG.

To the best of our knowledge, there is currently no literature review that provides a comprehensive in-depth content and SWOT analysis of school bus electrification. Therefore, the primary contribution of this review article is to outline the significant progress towards the national goal of achieving zero-emission school buses. This is accomplished by highlighting existing research on the current state of adoption, challenges of mass electrification, technological advancements in buses and charging infrastructure, bus deployment models, and the role of supportive policies in reaching these goals. Another key contribution is facilitating accelerated electrification by offering relevant information to stakeholders, including school districts, utility providers, school bus operators, and policymakers. Additionally, our analysis includes a SWOT analysis and identifies future research areas that could help achieve the widespread commercialization of zero-emission school buses.

This review article is organized as follows: The introduction provides a brief overview of the needs and challenges associated with the electrification of school transportation. The methodology section describes the methods adopted for this review, including the document search process, a flowchart, and a brief explanation of SWOT analysis. The content analysis and results sections discuss the current state of school bus adoption in the US, types of funding available, considerations for equity, and modalities of ESB deployment. This section also explains the critical role of policy in supporting school bus electrification. The SWOT analysis section outlines the strengths, weaknesses, opportunities, and threats related to school transportation electrification. The subsequent sections present future research areas aimed at accelerating school transportation electrification. Finally, this article concludes with a summary of the study's findings.

## 2. Methodology:

By utilizing VOS Viewer software for bibliometric analysis, we generated a thematic cluster map from the key terms found in Scopus within the "transport electrification" literature. The size of each circle on the map is proportional to the frequency of the keyword's occurrence, and the spatial proximity between terms indicates the strength of their association; the closer they are, the stronger their relationship [23]. The resulting thematic cluster map is a pivotal aspect of our review, as it illustrates the interconnected nature of electrification research topics, enabling us to visualize overlapping areas of interest and intersections between different research streams. However, we noted a rare or no inclusion of terms related to school bus electrification, ESB policies, and adoption methodologies in this cluster map, suggesting that this is a critical area for future research. In addition, these topics are predominantly discussed in white papers or grey literature, and are therefore not mentioned as keywords. This indicates the need for more extensive research documentation in the ESB to enrich the academic dialogue and support the advancement of school bus electrification.

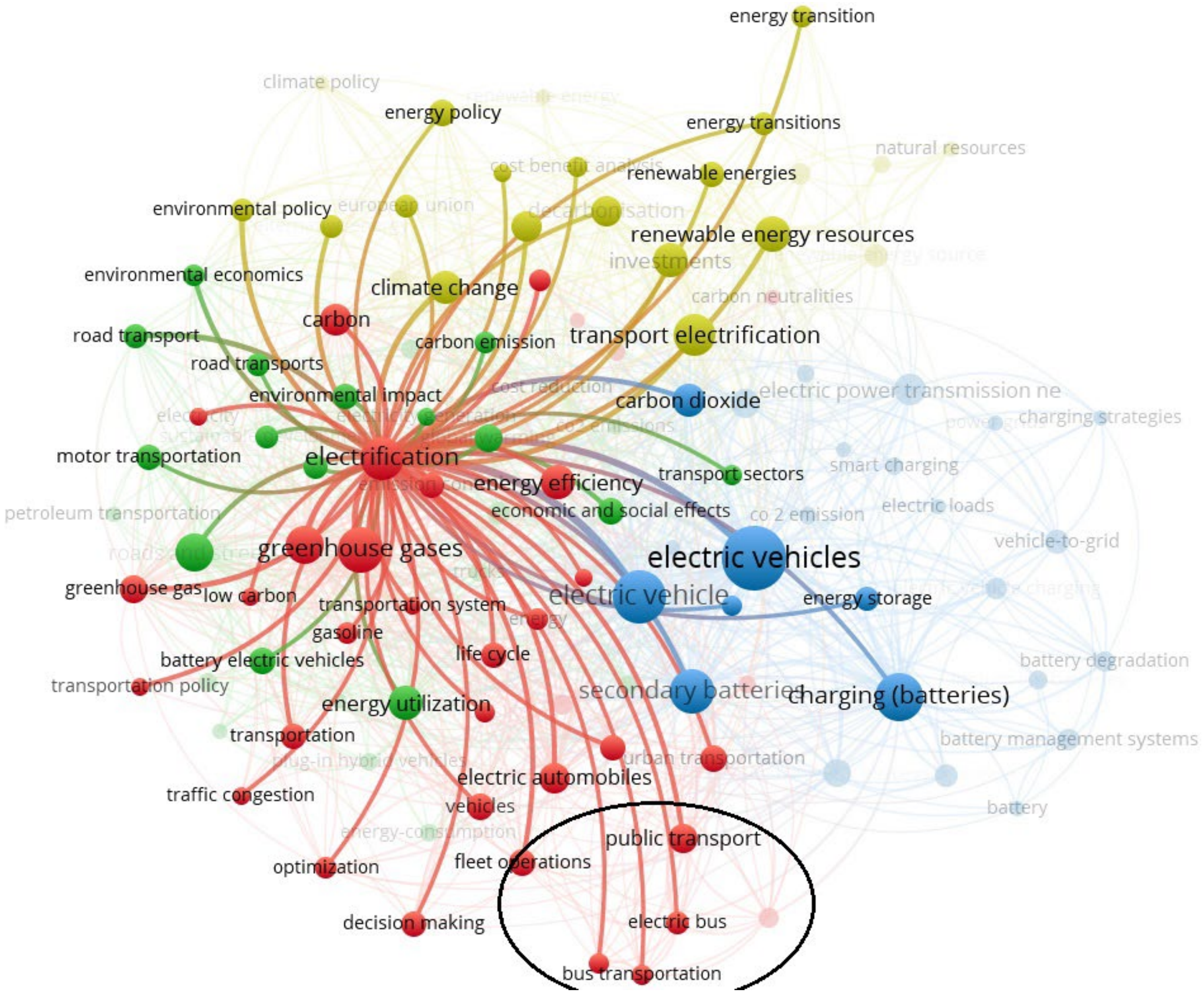

*Figure 2: Keywords Thematic Cluster Map Analysis to Synthesize the Research Gaps*

To underpin our review with a robust foundation of diverse and relevant sources, we conducted an extensive search across several acclaimed databases, including Google Scholar, IEEE Xplore, and ScienceDirect, and used the search process as many review papers do [24], [25]. The primary search was guided by carefully selected keywords and phrases specifically chosen to capture the breadth and depth of the topic. These included combinations such as "electric school buses" AND "zero emission" AND "government incentives", "school bus electrification", "fleet electrification school bus operators", "school bus fleets energy footprints", "transportation electrification technology and market assessment", "electric school buses bus operators", "bus electrification school districts", "electric school bus adoption" AND ("incentives" OR "subsidies" OR "grants" OR "regulations")", "electric school buses clean energy future", "electrification of public transport", "electric school bus series", "electric school buses" AND "electrification" AND "Policy" AND "US" and several others that varied slightly to widen the scope of our literature capture. This initial step helped identify a preliminary set of documents, which were further refined through a meticulous process of abstract reading, content skimming, and duplicate removal [23], [26].

The aim was to distill a core collection of literature that was both comprehensive and pertinent to the themes of electric school bus adoption, policy implications, and technological landscapes. From the initial findings, the documents were categorized and scrutinized, resulting in a final selection that included over 40 peer-reviewed journal articles, 30 reports and book chapters, three theses, and more than 30 white papers, policy papers, and official website articles. This compilation was rich in content, ranging from case studies on greenhouse gas (GHG) emissions and assessments of transport decarbonization to comparisons of emission metrics between diesel and electric buses, progress reports on ESB growth, and statutory policies set by states in the US on ESB adoption.

Our analytical methodology not only focused on summarizing these studies but also on integrating insights from the grey literature and real-time data sourced from reputable online platforms [27]. This approach was crucial, especially given the rapid evolution of school bus electrification post-2012. Regulatory updates and the latest governmental strategies were also included by accessing the most recent publications from official government, non-governmental websites, and blogs. By weaving together these various strands of literature, our review methodically maps the current state of ESB adoption, delves into the effectiveness of deployment models, and evaluates the environmental and policy-driven facets of the electrification movement. Through this comprehensive literature assessment, we aim to provide a holistic view of the advancements in school bus electrification and their implications for future sustainable transport initiatives. The distribution chart of the resources used is shown in the figure below.

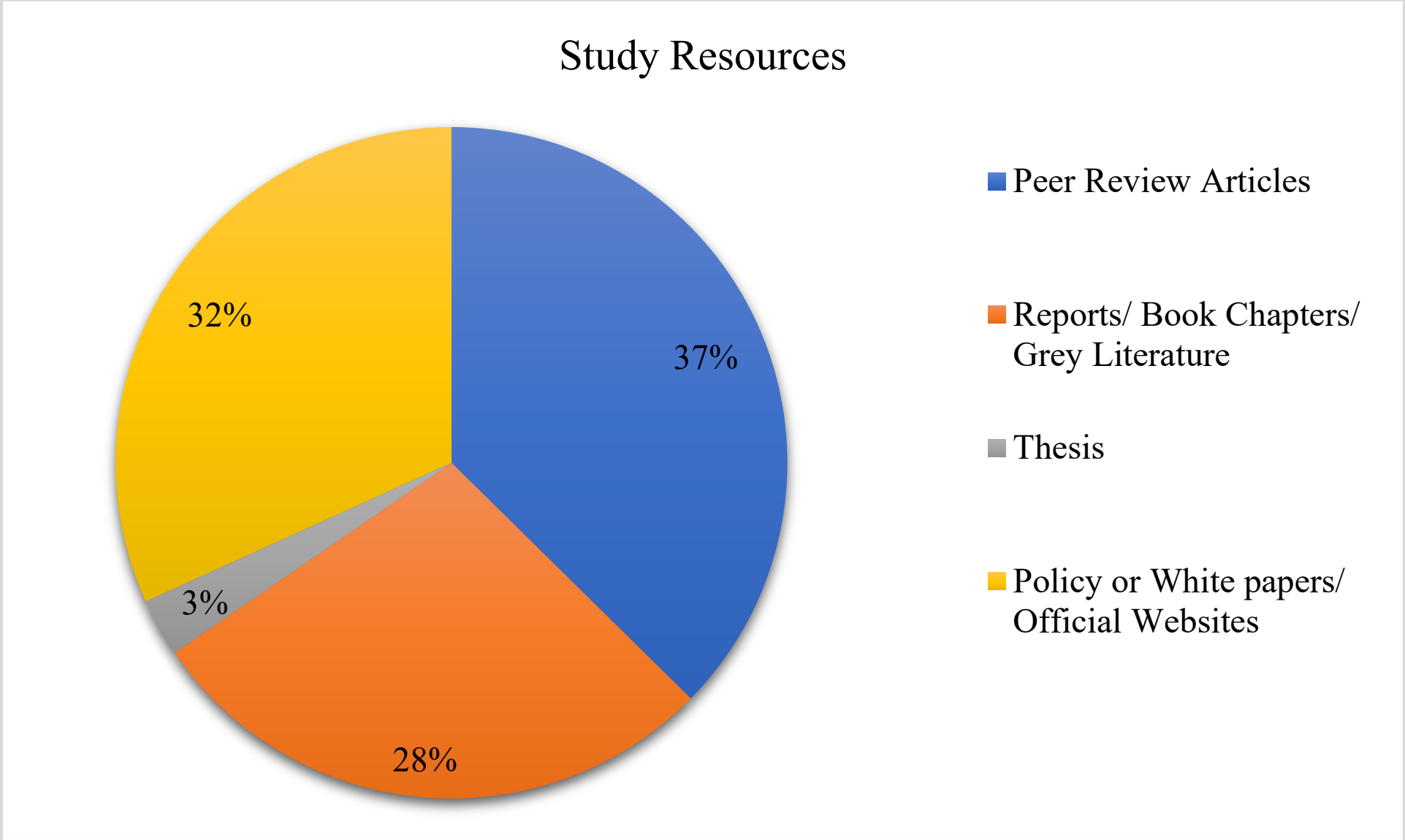

*Figure 3: Resources for our Study*

To visualize the key themes and vocabulary from our research documents, we employed a systematic method to create a word cloud. We used the Natural Language Toolkit (nltk), a Python package, which is a powerful tool for natural language processing (NLP). The size of each word in the word cloud map represents its frequency or significance in the source documents. This representation offers a quick, intuitive insight into the predominant concepts and terms within the field of school bus electrification. A world cloud map is shown in figure below.

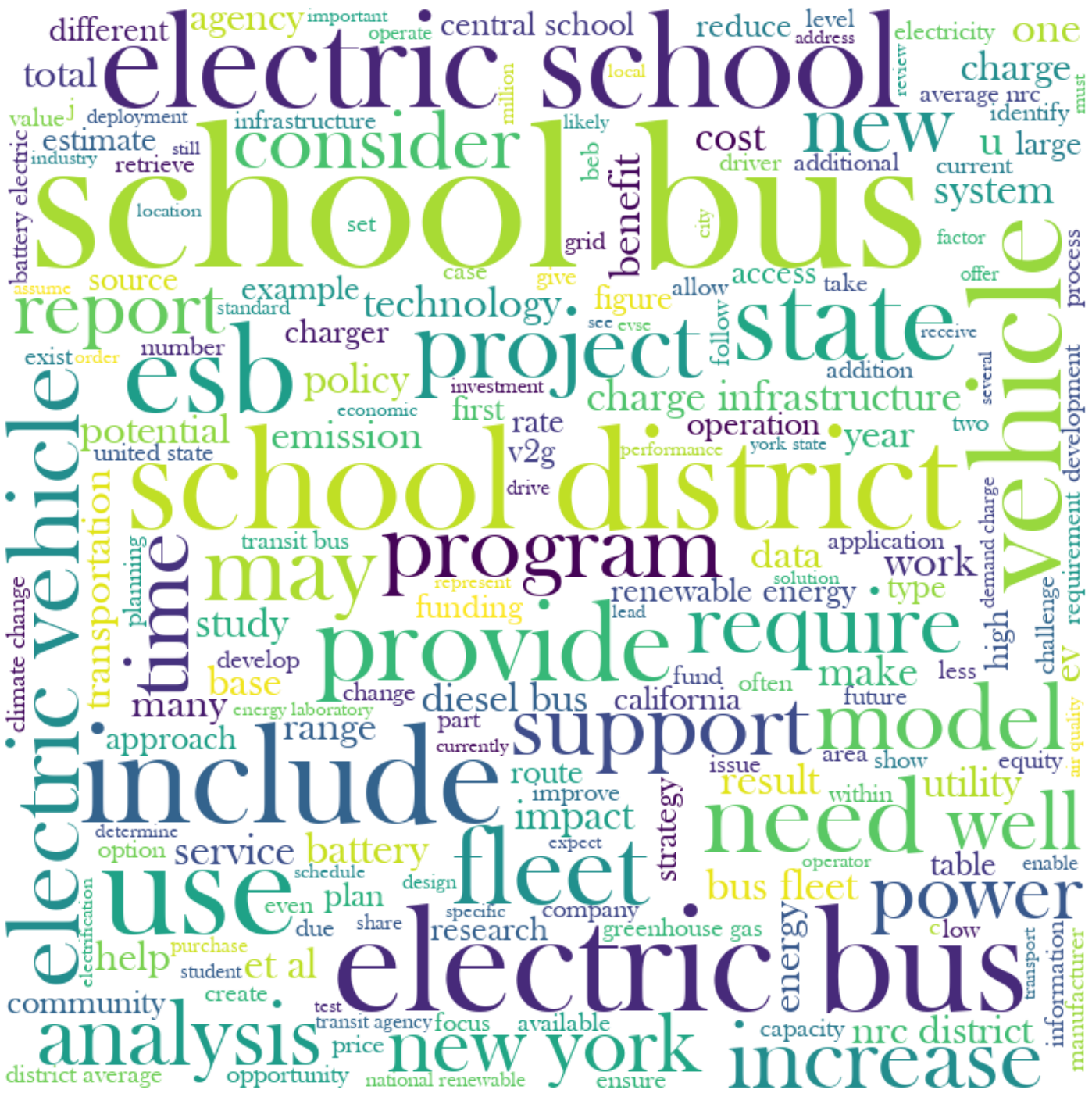

*Figure 4: Documents Word Cloud Result for Our Review Study*

After evaluating the selected documents, we conducted an in-depth content analysis of them. Content analysis is a research method that identifies patterns in recorded communication by systematically collecting data from texts, which can be written, oral, or visual [28] [29]. In our study, we utilized qualitative content analysis to systematically explore themes, patterns, and meanings within the topic of school transport electrification. This approach provided valuable insights by enabling a thorough examination of the data.

SWOT analysis, a well-established strategic planning tool used to identify Strengths, Weaknesses, Opportunities, and Threats, has been a staple in research for over six decades [30]. In this study, we applied SWOT analysis to deepen our understanding of the current state of ESB adoption, associated policies, and the direction of future research on large-scale school bus electrification. This involved a detailed examination of the relevant content and organization of this information into the appropriate sections of the SWOT analysis framework. We prioritized the most critical points by maintaining an objective and data-driven approach. Ultimately, we synthesized our findings into a SWOT analysis conclusion matrix, which highlights the strengths and weaknesses of ESB electrification, the opportunities it presents, and the potential threats to its widespread adoption. This methodological approach has been effectively used in various transportation-related review papers [31], [32], [33], underscoring its relevance and utility in our study. The methodological flowchart of our study is shown in figure below.

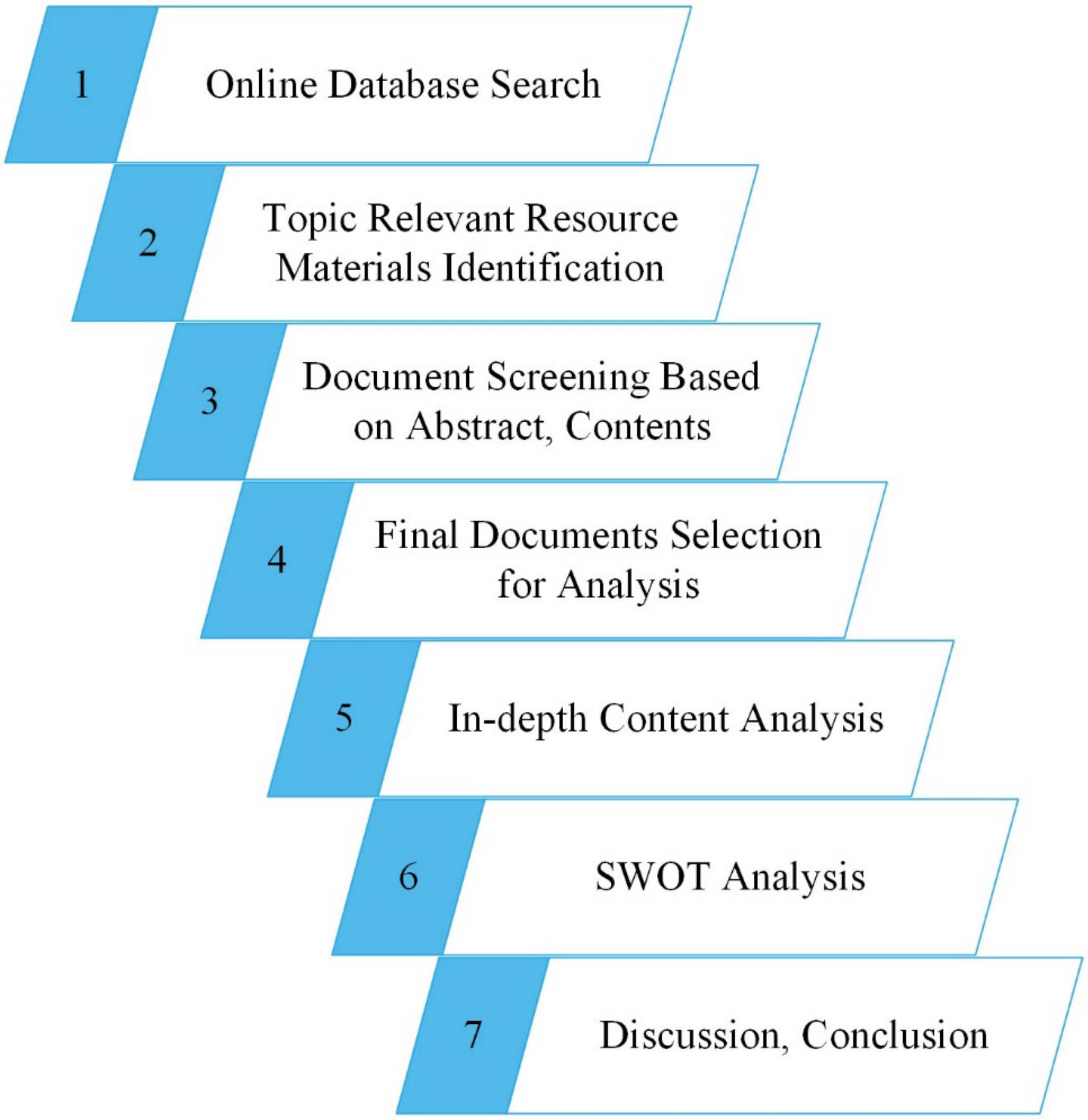

*Figure 5: Methodological Flowchart for Review Study*

## 3. Content Analysis and Results:

### 3.1. Need for School Bus Electrification:

School transportation plays a crucial role in ensuring the safe and timely movement of millions of students to and from educational facilities daily in the United States. The U.S. school bus fleet size is more than two and a half times larger than the combined number of vehicles in all other forms of mass transportation, making it the largest public transportation fleet in the nation [34], [35]. Traditionally, the iconic yellow school buses, which are a ubiquitous sight across the United States, have been reliant on diesel engines. However, this dependence on diesel technology has raised significant environmental and health concerns due to the emissions of harmful pollutants such as NOx, particulate matter, and volatile organic compounds. These emissions compromise air quality both inside the buses and throughout the communities they serve, impacting student health and contributing to broader environmental pollution. The particulate concentration on school buses is more than double the roadway concentration and fourfold that of outdoor concentration [36]. These emissions not only contribute to climate change but also generate air pollutants that are harmful to children's health, particularly fine particulate matter [37]. Children are more prone to air pollution than others because of their developing respiratory systems, which highlights the urgent need for cleaner school transportation [38]. Health issues, academic achievement, and burning environmental concerns have pressured the school bus program towards zero emissions.

ESBs offer a solution by reducing emissions, improving air quality and community health [3]. A comprehensive life cycle analysis revealed that BEBs can reduce global warming emissions by as much as 70%, even when the emissions from electricity generation are considered. This makes electric buses far superior in terms of environmental impact, with their effectiveness varying across regions; for instance, they are between 1.4 to 7.7 times more efficient than diesel buses in reducing emissions [39]. Operating on the national electricity mix, an electric bus emits approximately 1,078 grams of $CO_2e$, less than by a natural gas bus at 2,364 grams $CO_2e$ and a diesel-hybrid bus emiting 2,212 grams $CO_2e$ per mile [6], [38], [39]. The emission contribution chart as per Koehler [40] is presented in the Figure 6. Moreover, a shift to electric school buses alone could lead to a reduction of approximately 8 megatons in annual GHGs, accounting for a 35% decrease in emissions from all U.S. buses [13]. These buses are economically advantageous due to their high fuel efficiency, achieving an impressive 20.87 miles per DGE, almost 60% more energy efficient than their diesel counterparts [14]. The NY state anticipates saving at least $1.5-$2.8 billion in avoided emission-related damages by electrifying its school buses [19], [41].

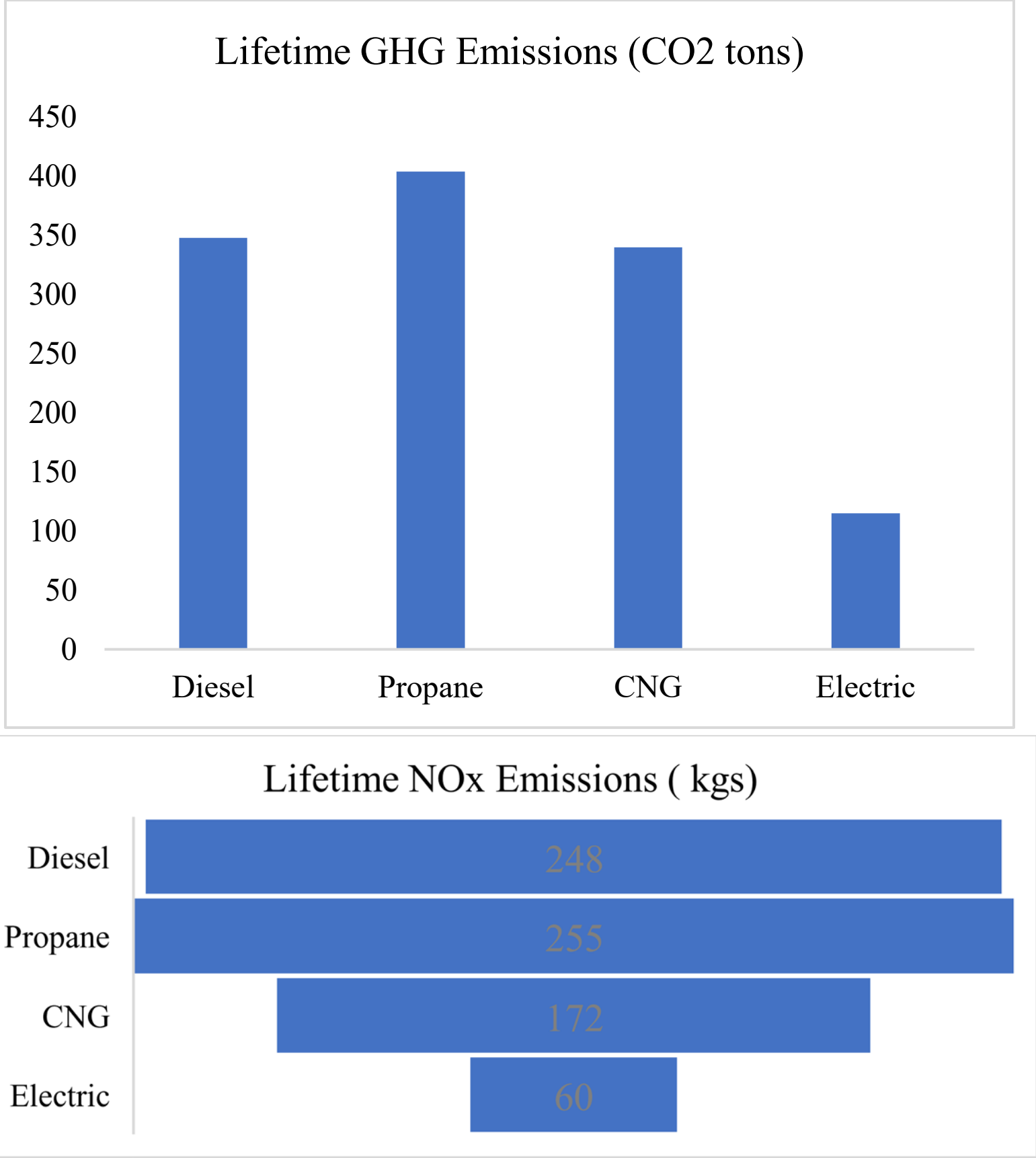

*Figure 6: Emission Contribution from Different Buses*

ESBs pose significant maintenance advantages over their diesel counterparts, where a typical diesel engine consists of approximately 2,000 moving parts, an electric bus motor has only about 20 [37]. EBs require less frequent attention to fluids; they need fewer oil changes and transmission fluid checks [42]. The overall cost-effectiveness of maintaining EBs is enhanced by lower labor costs, as technicians spend less time on routine maintenance tasks. Although the initial investment in electric buses may be higher, the long-term savings from reduced maintenance expenses can substantially offset these upfront costs, making ESBs a financially and operationally efficient choice in terms of TCO [15]. Furthermore, the health benefits of electric buses are equally compelling. This proactive change prioritizes the well-being of those most frequently exposed to bus emissions [6], [43], [44], [45].

### 3.2. Challenges of School Bus Electrification:

The transition to zero-emission school transportation faces several significant barriers. One of the primary challenges is the high upfront costs [17]. Although they offer cost savings over their operational lifespan, the initial outlay for purchase can be a substantial financial hurdle for schools. Additionally, the shift to zero-emission buses involves navigating a landscape filled with technological and policy uncertainties. Frequent changes in regulations, variability of incentives, and a lack of established infrastructure can create a complex environment that may deter schools from making the transition. Moreover, the infrastructure required to support electric buses, such as charging stations at schools and bus depots, is often inadequate. Another bureaucratic obstacle is the often-prolonged approval processes within school districts, which can significantly slow down the procurement of electric buses. Furthermore, the installation of the necessary charging equipment frequently requires upgrades to the existing electrical supply systems at educational facilities, adding another layer of expense and complexity. As per the graphic provided in Figure 1, we are presenting the challenges for ESBs and their progress as follows:

#### 3.2.1. High Upfront Costs:

ESBs come with higher initial costs compared to their diesel counterparts. For instance, Type A BEB with a student capacity of 10-30 are priced around $250,000 each, significantly more than the $50,000 to $65,000 range for diesel buses. Similarly, larger Type C with student capacity 54-90 or Type D with capacity 72-90 students cost between $320,000 and $440,000, while their diesel versions are around $100,000 [46]. Despite these steep upfront costs, electric buses present considerable operational savings, which are primarily in fuel and maintenance. Over their operational lifespan, electric buses can save between $4,000 and $11,000 per bus annually compared to diesel buses [37]. Market experts project that by the end of this decade, the total cost of ownership for electric buses, including both purchase and operational expenses, will reach parity with diesel buses excluding subsidies.

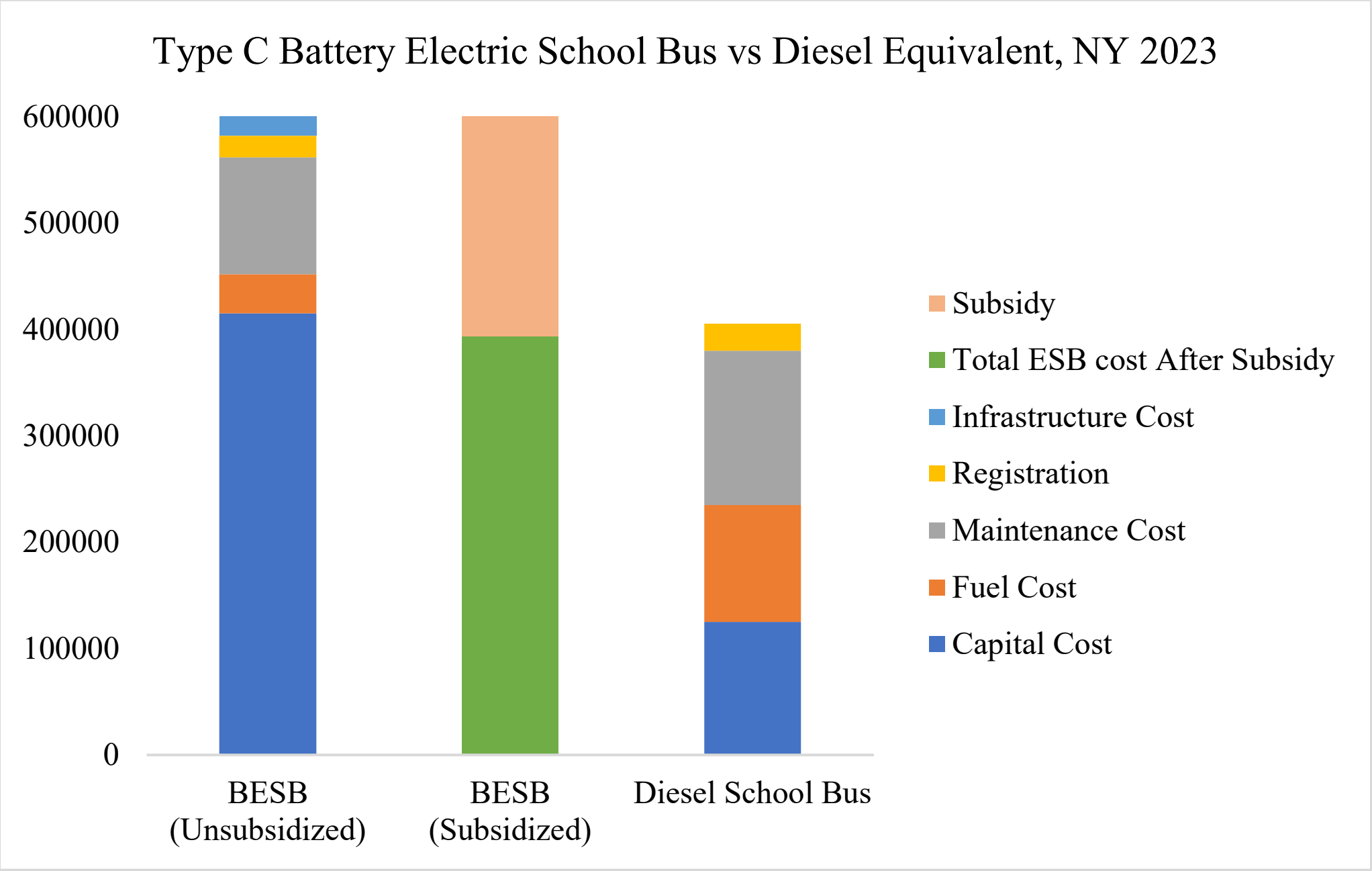

*Figure 7: TCO of ESB and Diesel Bus, NY, 2023 (data adapted from [19])*

The figure highlights the TCO of a Type C BESB, which is expected to run 60 miles per day and 200 days a year. The increased capital and infrastructure installation costs make the total bus cost high $608,000. Whereas New York state has incentives of 215,000, which includes $200,000 for the ESB purchase and $15,000 for the charger. With subsidy consideration, the purchase of ESB seems slightly lower TCO than the diesel counterpart in New York State. This is strong evidence to electrify newly purchased buses. In addition, the NY state ESB roadmap [19] forecasted that the price of ICE bus and BESB to be equal in 2027, even without the incentives or funding support.

Research by NYSERDA [19] indicates that the operational cost per mile for ESBs is 52% lower than that of ICEBs. They considered ESB charging efficiency 2.54 kWh/mile with the national electricity grid rate as mentioned. Moreover, the TCO for ESBs is optimized when they cover more vehicle miles, suggesting that electrifying longer or rural school routes can further reduce TCO. This strategy offers dual advantages: it not only lowers the TCO by utilizing ESBs for higher Vehicle Miles Traveled (VMT) but also prioritizes rural school districts, addressing equity concerns in the adoption of ESBs. Pacific Gas and Electric Company (PG&E) collaborated with Olivine and Liberty PlugIns on a pilot project to equip the Pittsburg Unified School District (PUSD). PG&E conducted an energy bill analysis and revealed that PUSD could save 20% per mile by transitioning from their current A-6 rate to the new BEV rate, optimizing charging to coincide with BEV TOU periods [47].

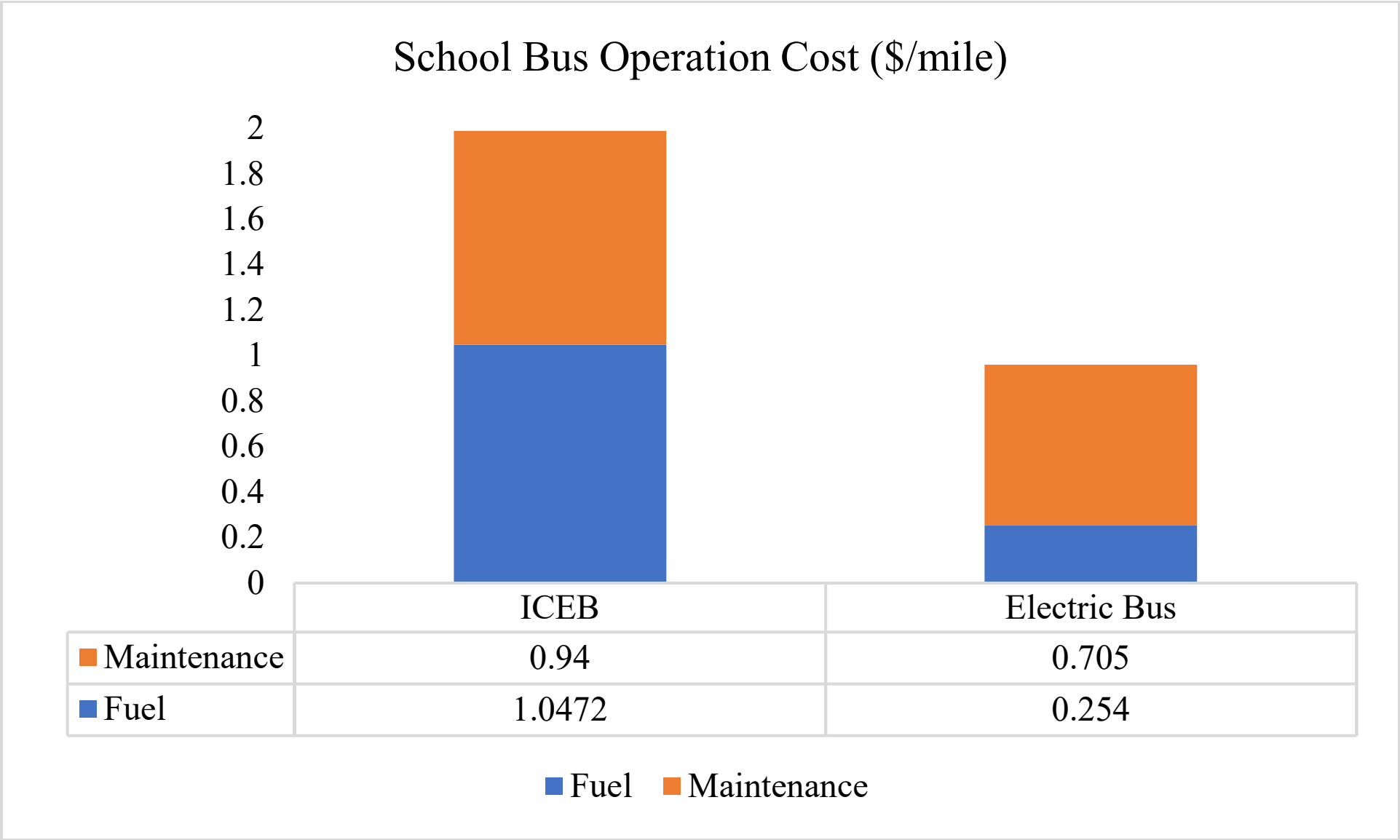

*Figure 8: Bus Operation Cost Comparison ICEB vs ESB (data gathered from [19], [47])*

Alternative bus ownership and operational strategies, such as "as-a-service" models, can mitigate initial costs and offer long-term cost stability. An emerging "repower" market is developing that transforms traditional fuel buses into electric ones, potentially reducing the cost of bus purchases by about 40% compared to new ESBs [19]. Thus, while the upfront investment is significant, the potential for long-term savings and the availability of supportive funding make electric school buses a smart choice for a sustainable future.

### 3.2.2. Lack of Charging Infrastructure:

The successful adoption of electric buses, particularly for school transportation, hinges significantly on the availability and reliability of charging infrastructure. The charger cost varies significantly from as low as $5,000-$15,000 for a Level 2 charger, which provides power limited to 19kW, to as high as $20,000-

$100,000 for a Level 3 fast charger providing power in the range of 50kW-350kW [19]. EV projection study by NREL suggested that there would be 33 million plug-in EVs by 2030, which requires cumulative nationwide capital investment of around $53- $127 billion in charging infrastructures, including private residential charging. To this end, $23.7 billion has already been announced for public charging infrastructure, where funds are supposed to be available by the end of 2030 [48]. An example of proactive infrastructure development can be observed in New York City. As 96% of district buses outside of New York City return to the depot during the evening and can charge overnight with a slow charger for reduced charging cost and reduced demand [19]. These infrastructure advancements are not just about supporting current needs but are also economically savvy in the long run. Therefore, investing in a robust and versatile charging infrastructure is essential for mitigating one of the main barriers to the wider adoption of electric school buses.

#### 3.2.3. Technology Uncertainty:

As battery technology rapidly advances, changes in battery chemistry, energy density, and charging capabilities continue to redefine the feasibility and efficiency of electric buses [43], [49]. There remains a degree of uncertainty around the actual range of these electric buses and the sufficiency of the existing charging infrastructure. School districts must critically evaluate whether electric buses can reliably cover all the required routes without the need for frequent recharges, which could disrupt daily school operations [18], [50], [51]. Average daily mileage per school bus in New York is under 100 miles, which is short enough and major routes could be electrified immediately [19]. However, in rural areas or other big cities length of school bus routes is long and needs careful consideration of charging. Duran and Walkowicz [52] reported that school route distances have up to 127.36 miles and 73.46 miles on average, with 154.46 miles in a 99.7% confidence interval. These examples signify the battery range and charging infrastructure planning importance for ease and accelerated school bus electrification. While it is well-known that electric buses generally require less maintenance due to fewer moving parts compared to diesel buses, questions about their long-term durability and the full spectrum of maintenance needs are still under investigation [18], [37], [42].

#### 3.2.4. ESBs and Charger Fire Safety:

An analysis by the federal National Transportation Safety Board (NTSB) on vehicle fires revealed that in 2022, for every 100,000 vehicles, 1,530 gasoline-powered vehicles experienced fire, while only 25 electric vehicles had similar incidents [53]. In the context of ESBs, as per Freehafer et al. [54], the United States had more than 2,277 units in operation as of early September 2023, with only a single reported bus fire caused by a computer component failure that notably did not spread to the battery [55]. Although ESB fires are much less frequent, they can pose different challenges, often reaching higher temperatures and demanding more resources for extinguishment. To mitigate risks, installation or enhancement of fire prevention systems is recommended to contain and prevent the spread of fire, should one occur [56], [57]. The National Fire Protection Association (NFPA) offers valuable resources to educate drivers and maintenance personnel on handling ESB emergencies, underscoring the importance of preparedness in the unlikely event of a fire [58]. To further minimize fire risks, fleet owners are advised to invest in high-quality, thoroughly tested ESBs and charging equipment, which reputable manufacturers consistently provide [56], [57], [59].

#### 3.2.5. Approval Delays and Procurement:

Because of several federal, state or other funding programs, there has been a rapid rise in ESB awards. However, school districts sometimes take a long time to approve electric bus purchases. Bureaucratic processes and decision-making can slow down the transition [60]. To smooth the bus procurement and adoption process, New York State has a policy that school districts have the option to acquire ESBs through the New York State Office of General Services (OGS) Procurement Services contract, which offers a pre-approved selection of bus makes and models. Alternatively, districts can conduct their own bidding process to customize bus specifications to meet their unique needs, or they can utilize third-party bus service contractors for the procurement of ESBs [19].

#### 3.2.6. Electrical supply upgrades and Grid Interaction Plan:

Installing electric bus chargers may necessitate upgrading the electrical supply in school facilities. These additional investments can be a barrier for some districts [61], [62]. Also, school buses are idle most of the time; therefore, schools may need a solid plan for using them in grid interaction. For bidirectional charging, we need charging technology, chargers that add a barrier. Due to charging grid stress, resilient grid infrastructure and peaking power plants also come into effect.

### 3.3. Current State of Electric School Bus Adoption

As of February 2024, electric school buses are committed in nearly all states. 47 states in the US have already delivered or started operation of ESBs, whereas 49 states except Wyoming, DC, four territories and several tribal nations have committed to ESBs. There is a total of 8,765 committed ESBs, which includes 3,482 awarded ESBs, 1,416 ordered ESBs and 3,867 delivered or operational ESBs in the US, where California state leads the way with 2,538 ESBs, where 343 are awarded, 598 ordered and 1,597 ESBs are delivered or operational [54]. The term committed consists of the buses either awarded/ funded to procure, procured, delivered or operational. With the updated dataset, it has been found that only 1.80% or 8,765 school buses are about to be electrified, whereas only 0.8% or 3,734 of the total school fleets are ESBs that are already on the road as of the end of 2022. A schematic diagram is shown in Figure 9. The electrified buses consist of both battery electric as well as plug-in hybrid electric school buses [22].

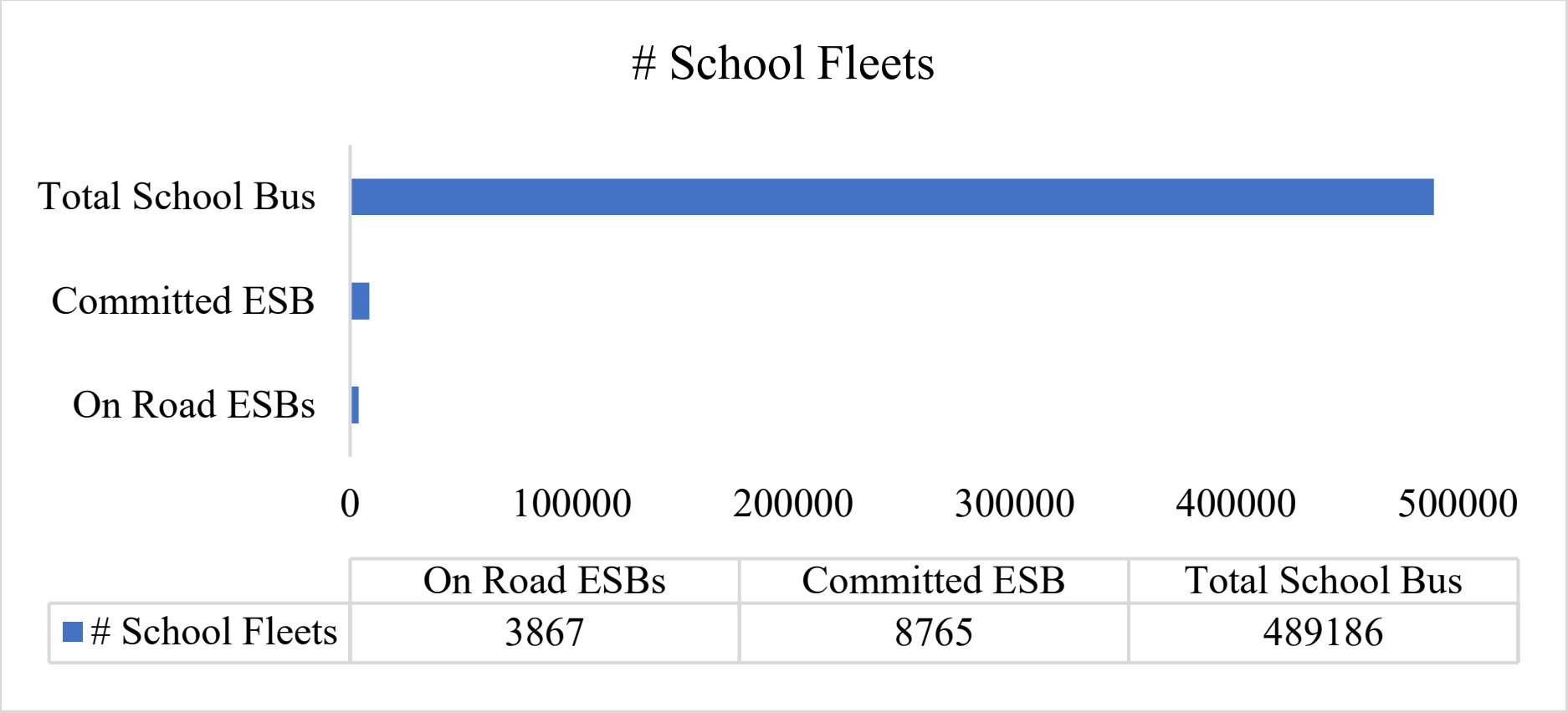

*Figure 9: School Bus Electrification Status*

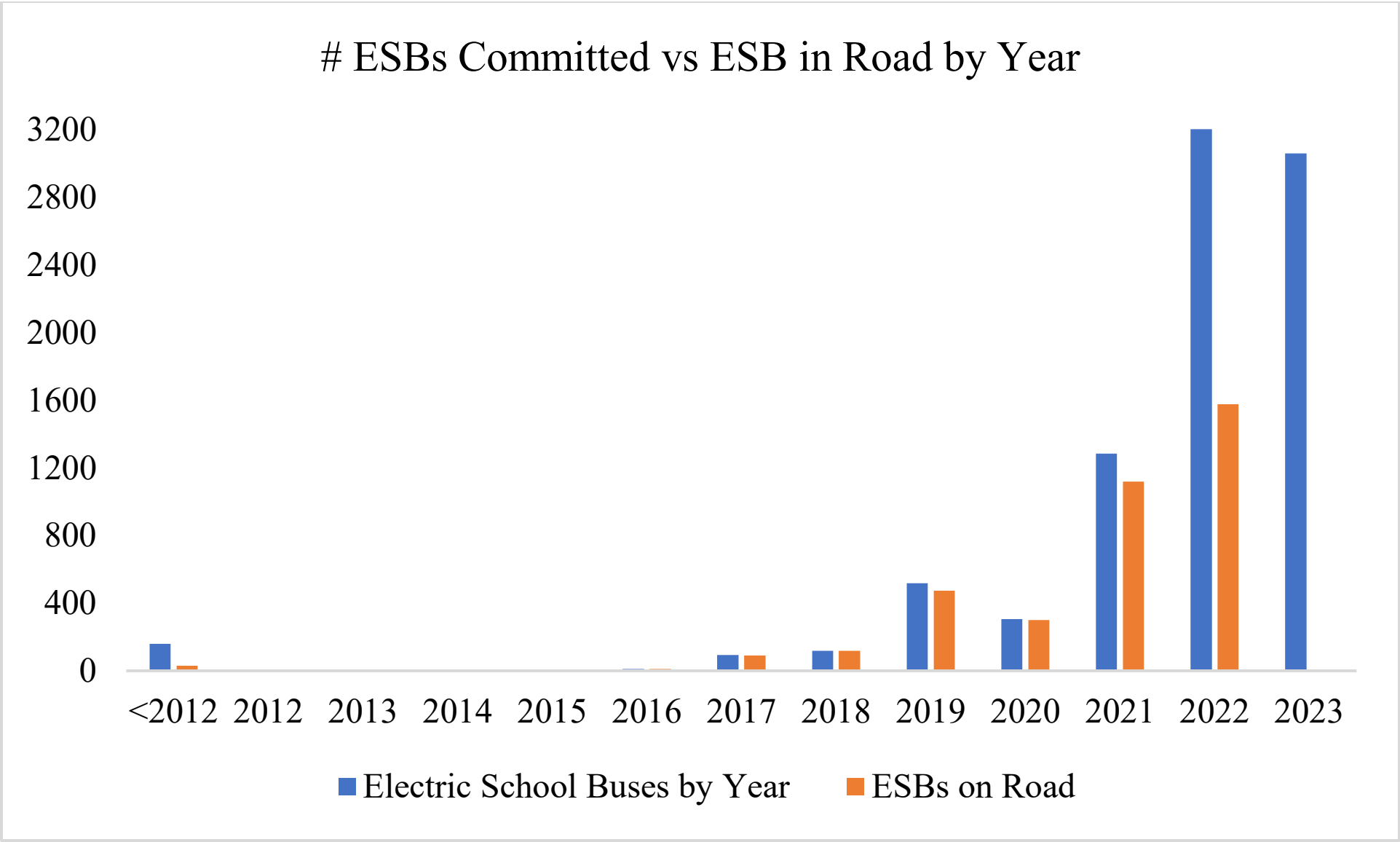

*Figure 10: Year-wise Committed Electric School Bus Adoption (adapted from data presented in [58])*

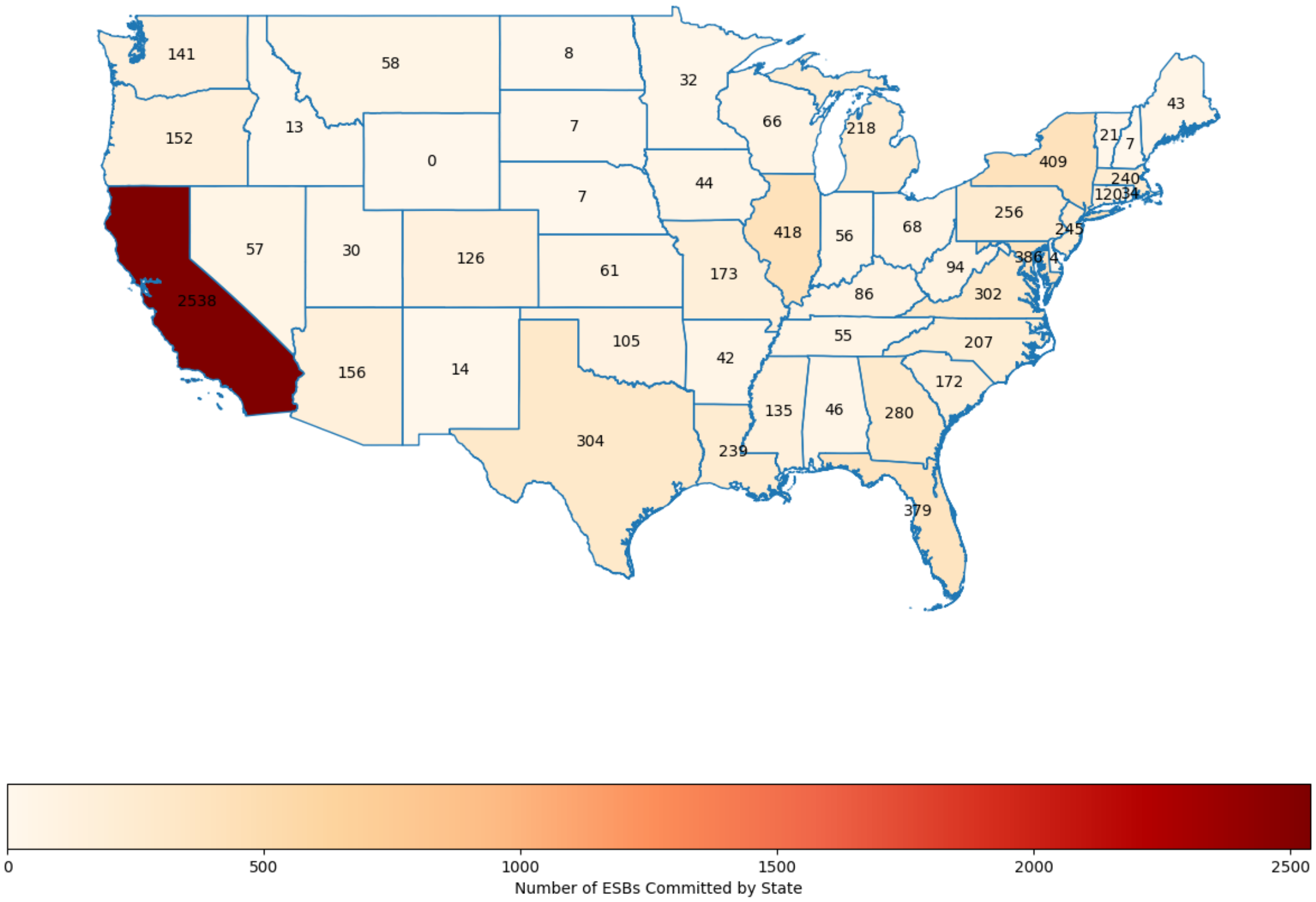

*Figure 11: Number of ESBs Committed by US States by March 2024*

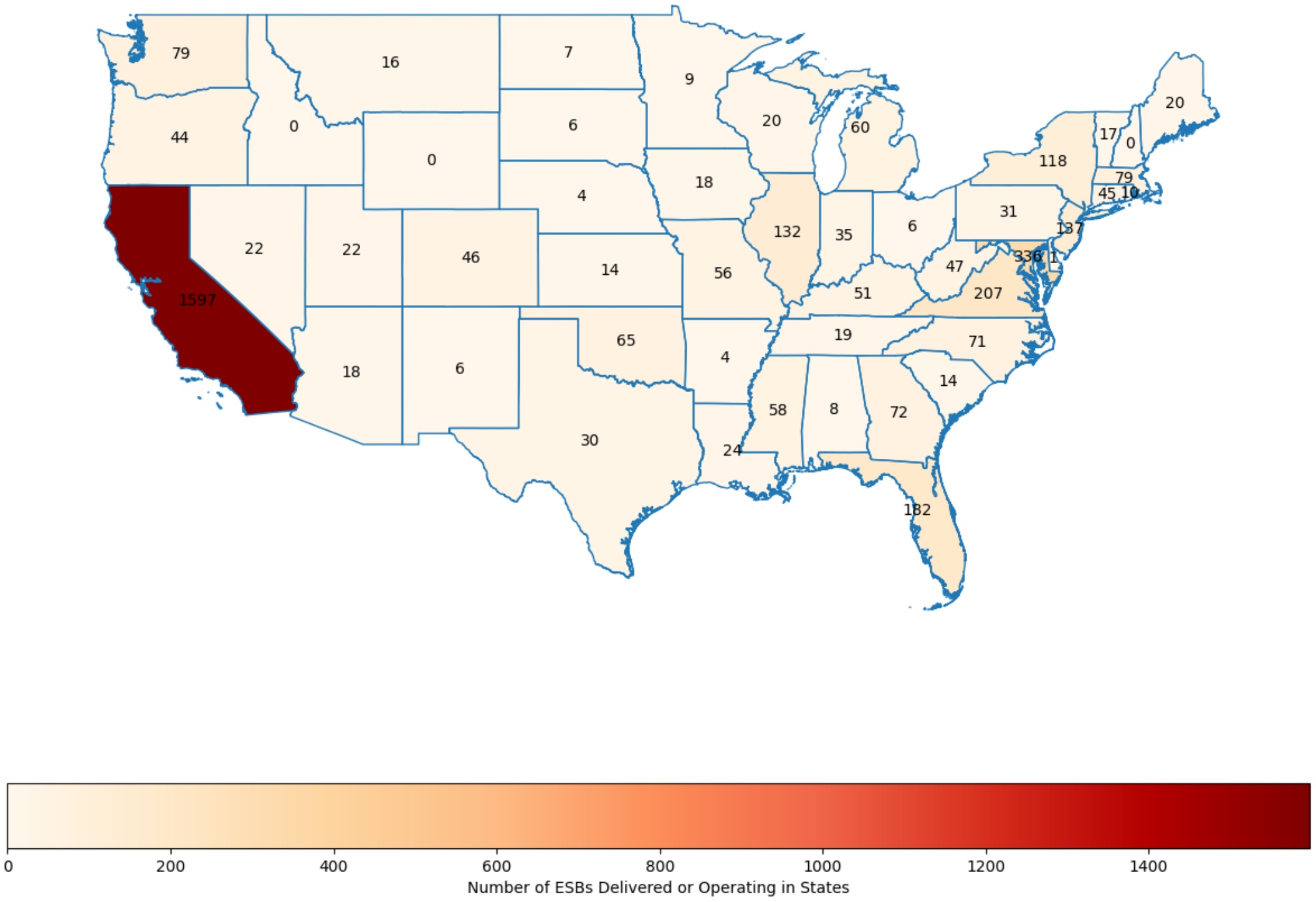

*Figure 12: Already Delivered or Operating ESBs in States by 2023*

It can be seen that the ESBs that are committed do not imply fully adopting. For each year, the committed ESBs are higher than the actual buses on the road. The data for before 2012 showed a drastic decrease in ESBs on the road. The reason behind this could be that the older buses may have been disposed of due to their useful life of those buses. It is also reported that there are 992 ESBs on order as of June 2023, which is around twice more as those ordered as of December 2022. This shows the steep up adoption of school buses in the US [54].

Among the top ten funding agencies or funding sources in the US, the EPA's Clean School Bus Rebate Program has played a significant role, awarding over $900 million for more than 2,300 electric school buses to 365 school districts in 2022. As per Lazer & Freehafer [22] and [63], the distribution of funding agencies on ESB adoption is shown in Table 1.

*Table 1: ESBs Funded through Major Provider Agencies and Programs in the US*

| SN | Agency Name | Funding Source/ Program Name | # ESBs Funded | Total |
|---|---|---|---|---|
| 1. | Environmental Protection Agency (EPA) | Clean School Bus Grant Program | 2,675 (awarded) | **5,154** |
| | | Clean School Bus Rebate Program | 2,339 | |
| | | Diesel Reduction Act | 140 | |
| 2. | California Air Resources Board (CARB) | California's HVIP | 1,172 | **1,413** |
| | | Rural School Bus Pilot Project | 143 | |
| | | Carl Moyer Program | 133 | |
| | | Community Air Protection Incentive Program | 108 | |
| 3. | Other Multiple Agencies (FL, MA, NJ) | Volkswagen Settlement (VW) | 712 | **899** |
| | | Regional Greenhouse Gas Initiative | 187 | |
| 4. | California Energy Commission (CEC) | School Bus Replacement Program | 228 | **228** |
| 5. | Dominion Energy (VA) | Electric School Bus Program | 59 | **59** |

Likewise, if we segregated the funding program by type of funder/agency/government. The data comes in order of the Federal government, state government, Volkswagen settlement, regional government, utility provider, private sector, cities and non-profit organizations. The data for committed ESBs of 5,086 led by the federal government to only two ESBs by cities and nonprofit organizations.

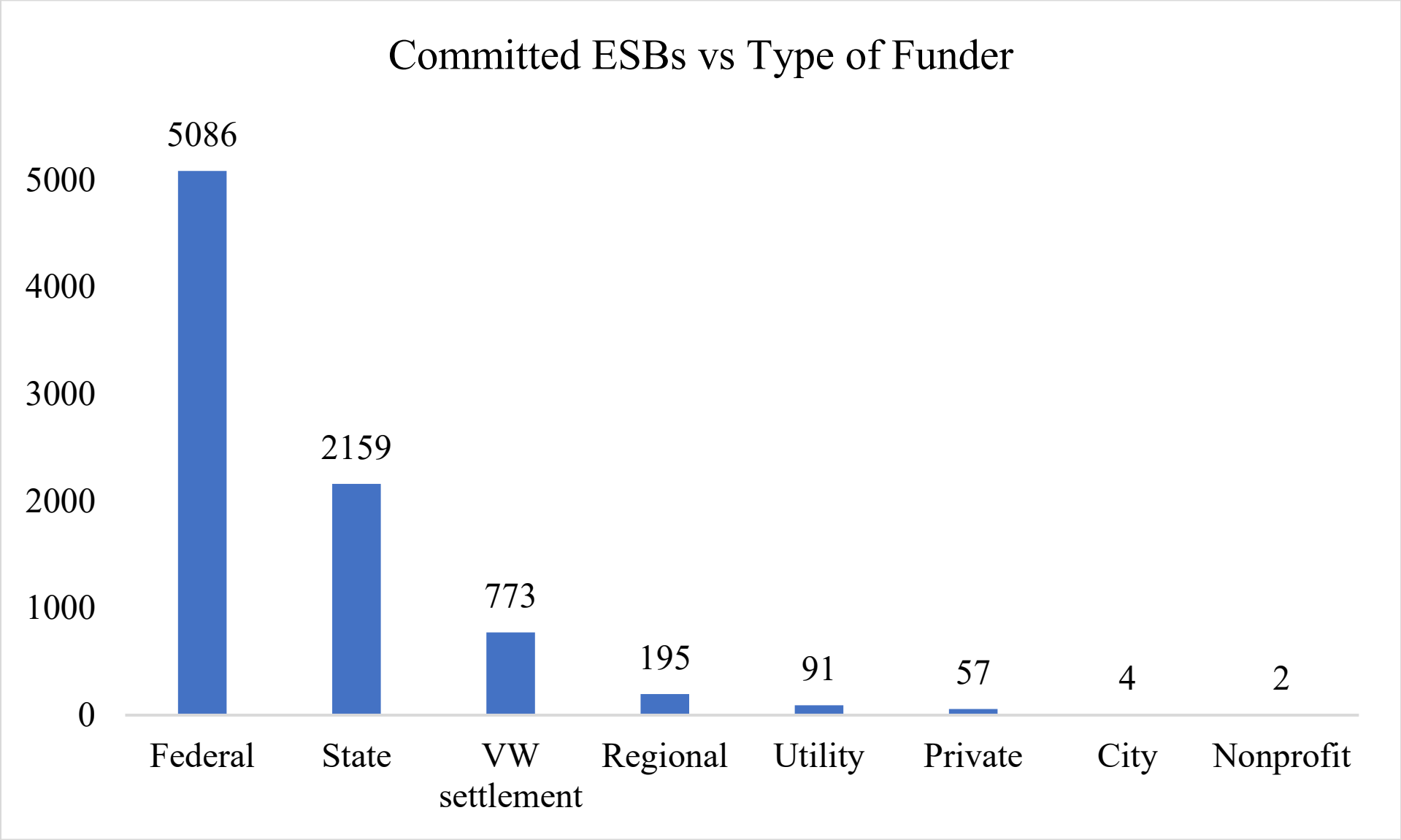

*Figure 13: ESBs by Different Types of Funder/ Agency*

Moreover, Lazer and Freehafer [22] summarize that Montgomery County Public Schools in Maryland not only has 86 ESBs in operation but also leads with a total of 326 committed ESBs. The Los Angeles Unified School District in California ranks second with a commitment to 253 ESBs, although it did not appear in the top five for current operation or delivery, indicating significant plans for bus electrification. New York City Public Schools in New York has committed to 118 ESBs, showing their intention to join the ranks of districts leading in sustainable transportation. Twin Rivers Unified School District in California is also prominent, with a current operation of 58 ESBs and a commitment to reach 84 ESBs, which demonstrates their ongoing dedication to transitioning from diesel to electric buses. Lastly, Troy Community Consolidated School District 30-C in Illinois is committed to 64 ESBs, a figure that places them in the top five for future electric school bus deployment despite not being listed in those currently in operation. In summary, they highlighted the leading school districts in the U.S. that are not only actively incorporating electric school buses into their fleets but are also making significant future commitments to expand their electric transportation systems.

### 3.4. Types of Funding and Financing ESBs and Their Current Scenario

The landscape of funding and financing for ESBs has become increasingly favorable to reduce the TCO and make electric school buses a more accessible option [18], [22], [54], [64], [65], [66]. **Grants:** This type of funding involves awards granted to applicants who meet specific criteria for a particular purpose, such as the EPA's Clean School Bus Program, which in 2023 awarded grants to eligible applicants for new electric school buses [67]. In addition, school districts will be compensated by the State Education Department via Transportation Aid, which will cover expenses up to 90% for some districts in New York, depending on their aid ratio [19]. Other grant programs include the Rural school bus pilot project (RSBPP) funded by CARB's low carbon transportation investment and air quality improvement program, CEC's School bus replacement program and Energy infrastructure incentives for Zero-emission commercial vehicles (EnergIIZE) program, CARB's Clean mobility schools pilot project and Community air protection program [68]. **Rebates:** These are reimbursements provided after the purchase of pre-approved equipment, with the EPA's Clean School Bus Program also offering rebates as a form of financial aid. The newly approved $500 million Environmental Bond Act in New York State offers financial incentives for school districts. **Vouchers:** Representing immediate credits that reduce the purchase price, vouchers like those from the NYSERDA's NY School Bus Incentive Program are given on a first-come, first-served basis to qualified applicants [3]. Also, there is a New York Truck Voucher Incentive Program (NYTVIP). Similarly, CARB funded the HVIP program, which was started in 2010 and $88.8 million has already been funded through this program specifically for school bus fleets in small air districts [68]. **Tax credits:** Tax-exempt entities such as school districts might benefit from tax credits, which can sometimes be received as direct payments. The Inflation Reduction Act of 2022 introduced tax credits for the purchase of new electric school buses and related charging infrastructure, such as the Qualified Commercial Clean Vehicles (45W) and the Alternative Fuel Refueling Property (30C) [19], [69].

When the full costs of electric bus procurement are not covered by these funding mechanisms, school districts can explore several financing opportunities: **Loans and Tax-Exempt Lease-Purchases:** These arrangements involve borrowing funds from public or private entities, with repayment terms typically set by the lender. **Bonds:** Commonly, municipal bonds issued by school districts finance educational infrastructure projects, with repayment often sourced from future tax revenues or other designated funds, usually requiring voter approval. **Operating Leases:** School districts may enter into operating lease contracts with original equipment manufacturers (OEMs) or dealers, which can include the option to purchase the bus at the end of the lease term. **Revenues:** ESBs have the potential to generate additional revenue through V2G charging by supplying excess power back to the electric grid during idle periods, such as midday and over summer breaks when electricity prices are highest. While V2G technology offers a promising revenue stream through utility demand response programs or New York State's Value of Distributed Energy Resources tariff, its financial viability remains uncertain [61]. Initial findings from a pilot in White Plains, NY, reveal that while revenues from selling energy can exceed charging costs, they do not always cover the expenses associated with V2G setup, such as upgraded charging equipment and potential battery replacements due to accelerated degradation from V2G use [70]. Through the combination of these diverse funding and financing options, school districts are better positioned to advance their plans for cleaner, more sustainable school transportation solutions [65].

### 3.5. Equity and School Bus Electrification:

Climate change and poor air quality impact falls disproportionately on historically disadvantaged communities. Shifting to cleaner buses is especially important for low-income students [71]. Across the U.S., 60 percent of low-income students ride the school bus, compared with 45 percent of other students [37]. Equity in ESB adoption is a crucial consideration to ensure that the benefits of clean transportation are distributed fairly across communities. Historically underserved regions, particularly low-income and disadvantaged neighborhoods, often bear a disproportionate burden of transportation-related air pollution and associated health risks. The adoption of ESBs in these areas can lead to significant improvements in air quality, directly impacting the health and well-being of the most vulnerable populations, including children. Prioritizing ESB deployment in these communities aligns with broader equity goals by rectifying historical disparities in environmental quality and health outcomes. Moreover, equity-focused ESB adoption strategies can foster inclusive participation in the green economy, create job opportunities, and ensure that all students, regardless of their socioeconomic status, have access to clean, safe, and reliable school transportation.

**Equity and Geographical Location:** The EPA's Clean School Bus Program, has taken an equitable approach, particularly favoring high-need, rural, tribal areas, and U.S. territories. This program has effectively distributed ESBs across different locales, such as rural, town, suburban, and urban, to closely match the national distribution of school districts. We found that all the school buses in the US are distributed 21% in urban, 33% in suburban, 15% in towns and 29% in rural areas. Whereas the distribution of ESBs has different figures, which account for 40% ESBs in urban, 32% suburban, 9% town and 16% in rural areas [54], [67]. This clearly shows that concentrated ESB adoption in urban areas, underscoring the program's inclusive reach.

**Equitable Adoption in Underserved and Low-income Population:** Currently, 62% of ESBs are committed to districts with high proportions of low-income households, a dramatic shift from earlier distributions favoring wealthier districts. Furthermore, 86% of ESBs are located in districts with higher populations of people of color, although race or ethnicity was not a specific criterion for prioritization [54], [65]. The California Energy Commission has launched the School Bus Replacement Program, dedicating over $94 million to help with electric alternatives, targeted at aiding disadvantaged and low-income communities. It underscores the state's commitment to promoting sustainable and equitable school transportation in vulnerable areas [72]. The Playbook for equitable electric school bus policy [64] acts as an exhaustive resource for state legislators, regulators, agencies, utilities, and advocates, providing detailed guidance on how to fairly transition their state's school bus fleet to electric power. As per this policy report, to ensure equity in deploying electric school buses, it is critical to engage communities and prioritize their introduction in low-income areas and communities of color most impacted by transportation pollution, while reserving 50% to 100% of state and utility funds, along with zero-cost financing and dedicated technical assistance, exclusively for these groups.

**Priority in Air Pollution Risk Areas:** In terms of environmental impact, school districts with the highest levels of particulate matter (PM2.5) and ozone, pollutants closely linked to diesel exhaust, have seen more substantial commitments for ESBs. Over two-thirds of ESBs are committed in districts with the highest PM2.5 levels, and approximately three-quarters are in areas with the most ozone pollution. The distribution among districts with varying adult asthma rates is more balanced, with 43% of ESBs in districts with the highest rates and 57% in those with lower rates. This suggests a need for continued efforts to direct ESBs to areas where air quality improvements would have the most significant health benefits [18], [54].

The World Resources Institute's ESB Initiative performed a baseline advocacy stakeholder analysis, engaging with 22 participants from 17 different organizations dedicated to equity and justice, such as environmental, disability, health, and tribal rights. The study revealed that most participants lack familiarity and involvement with national or local efforts to electrify school buses and do not possess sufficient knowledge about recent initiatives like the Infrastructure Investment and Jobs Act or the EPA's Clean School Bus Program [20]. This gap underscores the necessity for an informed and equitable push towards school bus electrification. Budzynski et al. [71] outlined four key dimensions of equity measures in school bus electrification, which encompass **procedural, recognition, distributive, and reparative equity**. Procedural equity ensures that communities most affected by electrification are actively involved in the planning and decision-making process. Recognition equity requires utilities to acknowledge and address historical disparities in energy and transportation access. Distributive equity seeks an equitable share of the benefits and burdens of electrification, ensuring fair investment distribution, like route selection and infrastructure upgrades. Reparative equity aims to rectify past and current injustices, such as improving air quality and providing targeted training for the maintenance and operation of new ESB technologies and infrastructure.

Supportive policies play a crucial role in advancing ESB adoption by creating an environment conducive to innovation and by reducing cost barriers. Utility companies, by advocating for such policies, can facilitate the transition for school bus operators through financial and technical support. Proactive policy-making can also enable these companies to enhance grid capacity in anticipation of increased electric demand from ESBs, aligning with state and federal initiatives aimed at promoting ESB usage.

*Table 2: Statutorily Enacted State Policies for ESB fleet transition*

| SN | State | ESB Transition requirements | Announced Funding |
|---|---|---|---|
| 1 | NY | 100% of new school buses ZEV by 2027; all school buses ZEV by 2035 | $ 500 million |
| 2 | Maryland | 100% of new school buses ZEV by 2025, if there is available federal or state funding | $ 200 million |
| 3 | Maine | 100% of new school buses ZEV by 2035 | Funding available under existing School Bus Purchase Program ($9.05 million in FY23 for all bus types) |
| 4 | Connecticut | 100% of all school buses electric by 2040 (or 2030 for buses operating in environmental justice communities) | $ 20 million |
| 5 | Delaware | 30% of new school buses electric by 2030. | - |
| 6 | California | 100% of new school buses ZEV by 2035, with a 10-year extension available to rural school districts | $ 1.5 billion |

### 3.6. ESB Deployment Models:

As per the National Association of Pupil Transportation (NAPT), over 30% of the national school bus fleets are privately operated and the remaining are operated through the school itself or students ride public transportation [73]. Student ride public transportation is the various transportation programs that allow them to ride public transit either free or at discounted rates. For example Washington Metropolitan Area Transit Authority (WMATA) has the Kids Ride Free Program and U-Pass [74], [75]. These alternative models provide schools with the option to outsource the transportation service. The ESB deployment models that are in widespread used in the US, as per Budzynski et al. [71] and Levinson and Curran [21] summarized and depicted in Table 3 below. The Transportation-as-a-Service (TaaS) model for electric school bus deployment offers an innovative solution to simplify the transition to electric buses while enhancing grid resiliency. In this model, a third party owns the school bus, charger, and infrastructure, managing operations and covering the high initial costs of electrification [18], [21]. Nuvve

Holding Corp's V2G Electric Vehicle Charging Hubs play a significant role in this model, offering holistic solutions that contribute to grid resilience [76].

*Table 3: ESB Deployment Models in the US*

| SN | ESB Deployment Model | Vehicle Owner & Maintenance | Charging & Infrastructure Owner | Operations Responsibility | Energy/ Fuel Manager (software) | Example School Districts |
|---|---|---|---|---|---|---|
| 1 | Traditional | SD | SD | SD | SD or Third party | Montgomery & Howard County Public Schools, MD<br>Twin Rivers Unified School District, CA<br>Los Angeles Unified School District, CA<br>Cambridge Public Schools & Acton-Boxborough Regional School District, MA |
| 2 | Lease | Third Party | SD | SD | SD or Third party | Valley Regional Transit (VRT), ID |
| 3 | Turnkey | Third Party | Third Party | SD | Third Party | Troy Community Consolidated School District, IL<br>Beverly Public Schools, MA |
| 4 | Transportation-as-a-Service (TaaS) | Third Party | Third Party | Third Party | Third Party | Select schools in CA, MD and FL |
| 5 | Charging-as-a-Service | SD | Third Party | SD | Third Party | Stockton Unified School District, CA<br>Fairfax County Public Schools, VA |

## 4. SWOT Analysis - School Bus Electrification:

### 4.1. Strengths:

The electrification of school buses stands out as a strategic move with wide-ranging strengths, encompassing environmental, health, and economic benefits. The environmental benefits are perhaps the most immediate, with electric school buses offering zero tailpipe emissions, which dramatically reduces the contribution to GHG and combats climate change while improving air quality and reducing noise pollution. The health implications are equally significant, particularly for children, as cleaner air translates directly to fewer respiratory problems and potentially improved academic performance due to better cognitive function in less polluted environments [5], [17]. The benefits of ESB transition are detailed in section 3.1. Moreover, the economic rationale for ESBs becomes clear when considering the TCO over time. Despite higher initial costs, electric buses tend to be more cost-effective in the long run [19]. In terms of operational efficiency, electric buses also showcase greater energy conversion from the grid to the wheels, leading to less energy wastage and lower operational expenses. Additionally, they contribute to addressing broader transport externalities, which include not only reducing traffic congestion and accidents but also mitigating the urban heat island effect often associated with traditional diesel-powered buses. These strengths collectively endorse the switch to electric school buses as an integral part of a sustainable, economically sensible transportation strategy that aligns with broader societal goals for a cleaner future.

### 4.2. Weaknesses:

The transition to ESBs comes with inherent weaknesses at present that pose substantial challenges to their accelerated adoption. The first challenge is the high initial cost of ESBs [77]. Additionally, the establishment of a sufficient charging infrastructure requires significant investment and logistical planning. The issue of range anxiety, where the fear that ESBs may not complete long routes on a single charge, adds complexity to their operation [78]. Technological shifts also necessitate driver and staff training, increasing transition costs. A detailed description of challenges to widespread ESB adoption is depicted in section 3.2. Expensive batteries are needed for ESBs, which face potential degradation and performance losses over time [79]. The supply chain for raw materials needed for these batteries, such as lithium, cobalt, and nickel, is also under strain, leading to potential volatility in pricing and availability. Furthermore, the increased power demands from widespread ESB charging could strain the electric grid, potentially causing imbalances and necessitating further investment in infrastructure [80]. Finally, the ever-evolving landscape of environmental policy adds a layer of uncertainty to the adoption process, with changes in legislation or incentives potentially affecting the viability of ESB initiatives. Addressing these weaknesses is crucial for a viable, long-term transition to electric school transportation.

### 4.3. Opportunities:

The electrification of school buses opens up a wealth of opportunities, greatly aided by the support of policies and financial incentives. Federal and state programs offering grants and rebates are pivotal in reducing the financial barriers [65]. In addition, the collaborative efforts between key stakeholders, including vehicle manufacturers, utility providers, and educational institutions, are crucial for fostering innovation and facilitating the sharing of resources and best practices [64]. Technological advancements further bolster the case for ESBs, with ongoing developments in efficiency and reliability, and the potential for integrating autonomous technologies promises to enhance safety and operational efficiency. The shift towards ESBs aligns seamlessly with the global imperative to decarbonize transportation and supportive energy policies. By adopting ESBs, districts are making a tangible commitment to environmental stewardship.

### 4.4. Threats:

The transition to ESBs is challenged by a variety of threats that could impede their widespread adoption [26]. Financial constraints are at the forefront. The volatility of energy prices, along with the need for physical space and resources to establish charging infrastructure, further complicates the economic feasibility of ESBs. Moreover, the sector is subject to technological and policy uncertainty, where rapid changes could disrupt the stability and consistent support for ESB initiatives. Competition from other clean transportation technologies also threatens the ESB market, potentially dividing attention and resources that could slow ESB growth [64]. The requirement for a resilient transportation system capable of overcoming technological malfunctions and external disruptions, such as natural disasters, adds another layer of complexity to the reliability of ESBs [81]. Cybersecurity poses a unique modern challenge, with the reliance on advanced technologies making ESBs potential targets for security breaches [82]. The conclusion of our SWOT analysis for accelerated ESB adoption and transport electrification is presented in Figure 14.

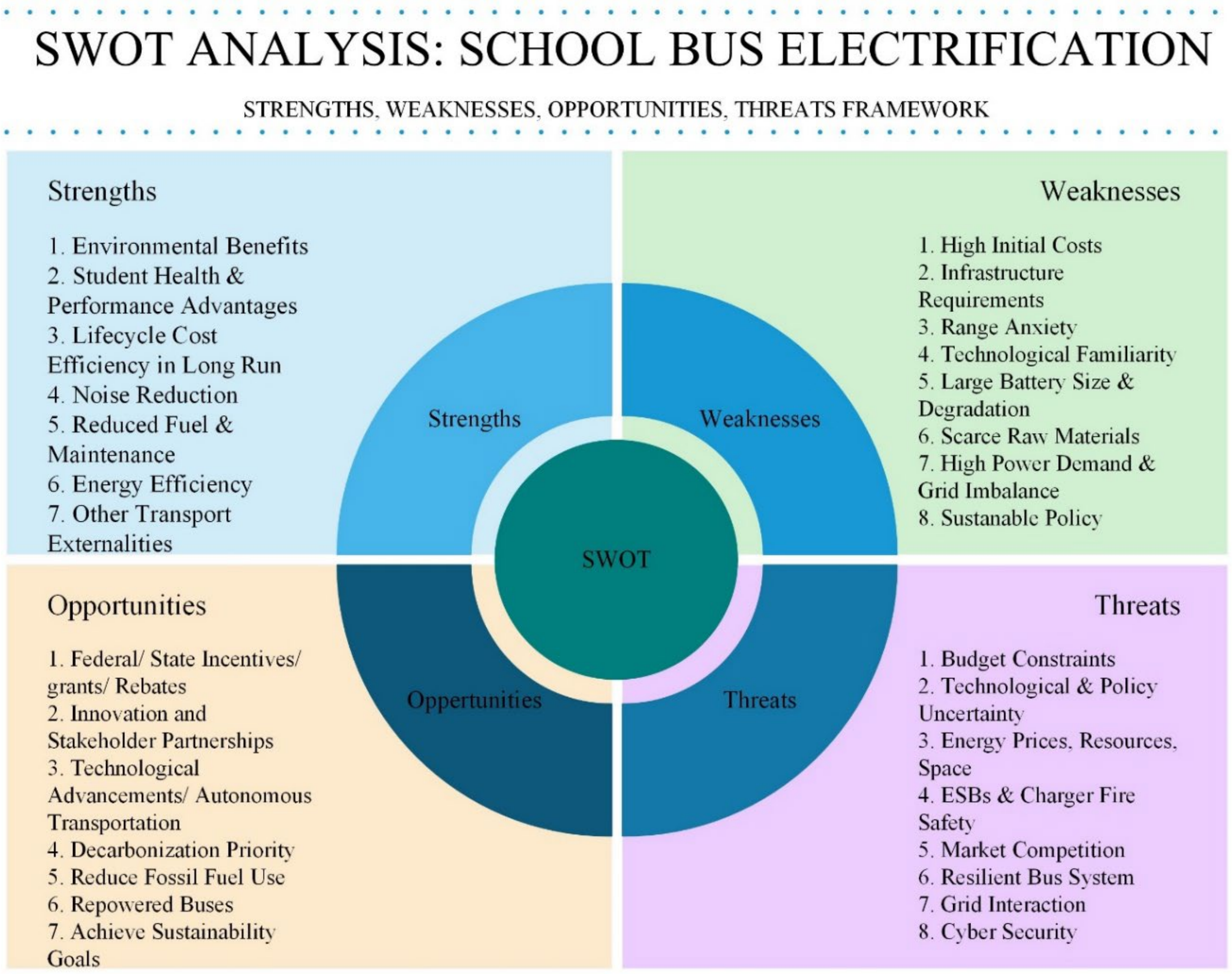

*Figure 14: SWOT Analysis Conclusions of School Bus Electrification*

## 5. Future Research Areas for Accelerated School Transport Electrification:

This section provides an analysis of critical gaps identified in the literature, which are expected to be key areas of focus, and discusses the current status of electrification and the future research required to advance the rapid electrification of school transportation in academic, industrial, and utility sectors. We highlight those unresolved issues in electric school bus operation and research, to be added to the existing literature.

**Dynamic Fleet Management Strategies:** Currently, the variable nature of ESB operations, characterized by stochastic behaviors, for instance, dynamic routing and optimization, remains underexplored. Dynamic route optimization reduces fleet size [83] for the same school transport requirements, which ultimately saves on the high purchasing cost of the school bus system, thereby contributing to rapid school bus electrification [84]. This gap presents a compelling case for the application of robust or stochastic optimization models in future research endeavors. External variables such as driving styles, elevation changes, the frequency of stops, types of routes, and environmental conditions like weather and traffic could significantly impact energy consumption [51], [78], [85]. These factors warrant further investigation to refine fleet management strategies. Future research should also explore the integration of varying battery capacities, diverse charging strategies, for instance, slow charging overnight and fast charging during the day or in between services, and the use of mixed bus fleets to achieve more efficient fleet management solutions [86]. As school bus networks expand on large scales, the number of decision variables and constraints in these models increases dramatically, often beyond the practical reach of exact algorithms, suggesting the metaheuristics for optimal or near-optimal solutions [87]. Furthermore, developing advanced energy consumption methodologies that allow for intra-day adjustments to electric bus charging, aligned with the actual energy demands of the vehicles, becomes essential [88].

**Impact on Grid and Strategic Interactions:** The development of innovative strategies to alleviate increased grid stress caused by mass school fleet charging becomes paramount. Among the potential solutions, smart coordinated charging strategies emerge as a critical area for investigation. These strategies should carefully consider the interactions between the school's electric bus systems and grid operations. Additionally, the dynamics of V2G interactions present intriguing possibilities for research. For instance, PTOs could leverage the special nature of the school bus system of having a predicted schedule of operations, almost 85-90% of idle time at depot or school and TOU prices to sell excess energy back to the grid during peak periods. This necessitates the development of real-time energy management systems capable of responding swiftly to these fluctuations. Furthermore, given their substantial battery storage capacity, EB fleets can function as portable energy storage systems (ESS), thereby enhancing grid resilience. Investigative studies focusing on these areas are crucial to supporting the accelerated and widespread adoption of ESBs and ensuring their effective integration into grid systems.

**Repowered or Retrofitted School Buses:** Repowered school buses, also known as "repowers," offer a cost-effective alternative to buying new ESBs. Essentially, repowers involve taking an old internal combustion engine (ICE) bus and retrofitting it with an electric powertrain, which includes removing the engine, transmission, and exhaust system and integrating the existing 12-volt system for auxiliary components like lights and compressors. These conversions can be done directly by a repowering company, which then sells the bus to the school district. This option is significantly cheaper than purchasing a new ESB, costing roughly the same as a new diesel bus at about $50,000 to $100,000 for a Type C bus, offering nearly 40% savings compared to new ESBs [89], [90]. Despite these advantages, the repower market is still small, with fewer than 10 buses currently operational in the U.S. as of April 2023. Constraints include the limited availability of suitable buses for conversion and the technical challenges of standardizing the installation across different bus models. However, the market is poised for growth, evidenced by initiatives like SEA Electric's partnership with Midwest Transit, aiming to repower 10,000 buses over five years [19], [91]. In NY State, repowers have historically qualified for funding through incentive programs like the NYTVIP, although wider financial support remains uncertain with current federal policies excluding repowers from certain grants.

**Advancements and Challenges in Autonomous Electric School Buses:** Autonomous driving represents a transformative force in the landscape of future mobility. While there is a wealth of research surrounding autonomous EVs, the specific area of autonomous electric buses (AEBs) has received comparatively less attention. Autonomous buses must navigate a unique set of challenges and requirements that differ substantially from those faced by conventional autonomous vehicles. These include specialized energy management and charging schedules, maintenance of battery health, ensuring inter-vehicle safety, optimizing for passenger comfort, and minimizing energy consumption [23], [92], [93]. Given the complexity of these requirements, future research must be directed towards addressing these specific challenges associated with AESBs. This should not only focus on the technological aspects but also consider the human factors involved, such as end-user acceptance.

**Enhancing Resilience of Bus Systems Against Extreme Events and Threats:** The resilience of school bus transportation systems during extreme events, natural disasters, and cyber-attacks is a critical area of concern. As climate change increases the frequency and severity of natural disasters, and as digital infrastructures become more integral to transportation systems, ensuring the robustness of bus networks against such disruptions is paramount [23], [81], [82]. In these cases, the bus system can work as a mobile power plant by using its battery stored energy to serve the critical infrastructures to remain operational, like hospitals, schools. In addition, the school has a large enough space to be used as a shelter during flooding, earthquake or extreme events. This involves developing comprehensive disaster response strategies, improving infrastructure to withstand environmental stresses, and enhancing cybersecurity measures to protect against digital threats (cyber-attacks on the grid system). Additionally, integrating advanced predictive analytics and real-time monitoring systems can significantly bolster the capability of bus systems to anticipate, respond to, and recover from these challenges effectively.

## 6. Conclusions:

Adoption state of ESBs in the US is encouraging year by year, although only 0.8% of ESBs are on the road to transport pupils, statistics show 1.8% of total buses are already committed through various funding programs implemented by various agencies like EPA, CARB, NYSERDA and CEC [22], [54]. Moreover, New York state anticipates the electric bus cost to be breakeven in TCO with diesel buses by 2027 [19]. The extent of school bus electrification has been controlled by the state of California as which holds around 29% share of total committed ESBs in the USA as of 2023. Such actions of aggressive school bus adoption could be achieved through integrated efforts of federal, state governments, utility providers and users. Furthermore, climate change and poor air quality impact falls disproportionately on historically disadvantaged communities, suggesting the more equitable school bus electrification policy. Budzynski et al. [71] outlined four key dimensions of equity measures in school bus electrification, which encompass procedural, recognition, distributive, and reparative equity. Moreover, many states have set out the statutory policies of having school transport be zero emission by 2035. Unlike the most common and traditional school bus deployment model, there have been changes in ESB deployment modality, such as leasing, turnkey and TaaS. Mass adoption involves a holistic approach, including infrastructure planning, training, end-of-use considerations, and collaborative efforts. The SWOT analysis underscored a multifaceted landscape, highlighting critical weaknesses, including high initial costs, infrastructure challenges, range anxiety, and the need for greater technology. On the opportunities front, the availability of federal and state incentives, technological advancements, bus retrofitting and the push for decarbonization offer promising avenues. Meanwhile, threats such as budgetary constraints, policy uncertainties, and the nascent concerns around ESB and charger fire safety present areas where strategic focus is necessary to mitigate risks.

We also provide analysis on critical gaps which are expected to be the key area of focus. Dynamic routing of electric school bus routes can reduce the required fleet size. To utilize the ideal time, predicted schedule and larger battery size of ESBs, the V2G interaction needs to be researched, which opens the avenue for revenue generation and demand charge management. Repowering or retrofitting school buses is significantly cheaper than purchasing a new ESB. While ESBs are growing, this is the perfect time to integrate autonomous driving in our electric school buses. Furthermore, electric school buses can work as a portable energy storage system to enhance the resilience of bus systems. The present work has limitations; for example, we used real-time data to support the current state of school bus electrification, which would update periodically. We provide a high-level review of school transport electrification, giving an overview of cost, technology and policy scenario and barriers. Each topic is worth a microscopic review. We used grey literature and/or white papers and policy papers to make our review article more realistic by supporting the newest trend in school bus electrification. Those white papers or policy papers need to be regularly examined and updated for adoption and policy status.

**Acknowledgement:**

This research was supported in part by the Center for Advancing Sustainability through Powered Infrastructure for Roadway Electrification (ASPIRE), a National Science Foundation (NSF) ERC, under award number EEC-1941524. Also, we would like to thank WRI for their series of updated datasets on school bus adoption that we used to inform the school bus adoption status.

**Author Contributions:**

**MBK** Conceptualization, Methodology, Validation, Formal analysis, Investigation, Data Curation, Writing - Original Draft, Writing - Review & Editing, Visualization; **ZS** Conceptualization, Methodology, Validation, Investigation, Resources, Writing - Review & Editing, Visualization, Supervision, Project administration, Funding acquisition. Both authors reviewed the results and approved the final version of the manuscript.

**Data Availability:**

Data would be available upon reasonable request to the authors.

**Declaration of Interests:**

None.